\documentclass[%
reprint, %
superscriptaddress,
showpacs,
nofootinbib,
amsmath,amssymb, %
aps, %
prd, %
twocolumn %
]{revtex4-1}
\usepackage[utf8]{inputenc}
\usepackage{hhline}
\usepackage[dvipsnames]{xcolor}
\usepackage{amsmath}
\usepackage{hyperref}
\usepackage{graphicx}
\usepackage{xcolor}
\usepackage{bm}
\usepackage{dcolumn}
\newcommand{\diff}{\mathrm{d}}

\newcommand{\be}{\begin{equation}}
\newcommand{\ee}{\end{equation}}
\newcommand{\ba}{\begin{eqnarray}}
\newcommand{\ea}{\end{eqnarray}}

\begin{document}


\title{Testing $F(Q)$ gravity with redshift space distortions}

\author{Bruno J. Barros$^{1}$, Tiago Barreiro$^{1,2}$, Tomi Koivisto$^{3,4}$ and Nelson J. Nunes}
\affiliation{Instituto de Astrof\'isica e Ci\^encias do Espa\c{c}o,\\ 
Faculdade de Ci\^encias da Universidade de Lisboa,  \\ Campo Grande, PT1749-016 
Lisboa, Portugal \\ 
$\,^{2}$Departamento de Matem\'atica,  ECEO, Universidade Lus\'ofona de Humanidades e 
Tecnologias, Campo Grande, 376,  1749-024 Lisboa, Portugal \\
$\,^3$Laboratory of Theoretical Physics, Institute of Physics,University of Tartu, W. Ostwaldi 1, 50411 Tartu, Estonia \\
$\,^4$National Institute of Chemical Physics and Biophysics, R\"avala pst. 10, 10143 Tallinn, Estonia}

\date{\today}

\begin{abstract}
A Bayesian statistical analysis using redshift space distortions data is performed to test a model of Symmetric Teleparallel Gravity where gravity is non-metrical. The cosmological background mimics a $\Lambda$CDM evolution
but differences arise in the perturbations. The linear matter fluctuations are numerically evolved and the study of the growth rate of structures is analysed in this cosmological setting. The best fit parameters reveal that the $\sigma_8$ tension between Planck and Large Scale Structure data can be alleviated within this framework.
\end{abstract}

\maketitle


\section{Introduction}\label{introduction}

In the last two decades, a plethora of ground and satellite based data of the cosmic microwave background radiation, galaxy distribution and supernova Ia brightness-redshift relation was gathered to convey the simple successful Lambda Cold Dark Matter cosmological model ($\Lambda$CDM). 
Assuming a $\Lambda$CDM model, however, a tension between CMB observations and redshift space distortion measurements on the amplitude of the matter power spectrum at the scale $8h^{-1}$Mpc,  ($\sigma_8$) is found \cite{Douspis:2018xlj,Battye:2014qga,Macaulay:2013swa,Nesseris:2017vor,Basilakos:2017rgc,Joudaki:2016mvz}. Planck values suggest a higher rate of clustering in comparison with late large-scale observations. Proposals to solve or alleviate this problem have been put forward assuming extensions of the standard model of cosmology, such as modified gravity \cite{Nersisyan:2017mgj,Amendola:2019fhc} and dynamical dark energy models \cite{Gomez-Valent:2018nib,Lambiase:2018ows,Barros:2018efl}. These have dynamical evolutions for the matter perturbations distinct  from the evolutions obtained with the standard $\Lambda$CDM, which consequently makes it possible to obtain values of $\sigma_8$ that are concordant. 

In this work we seek yet another explanation for the discrepancy as we study the evolution of matter perturbations in modified $F(Q)$ gravity \cite{Lazkoz:2019sjl,Jimenez:2019ovq,Lu:2019hra}, the nonlinear generalisation of the improved version of General Relativity \cite{BeltranJimenez:2017tkd} on a flat and torsion-free spacetime continuum known as Symmetric Teleparallel Gravity (STG) \cite{Nester:1998mp,Adak:2008gd,Adak:2018vzk}. There is a number of recent studies of modified STG models and their applications in cosmology \cite{Dialektopoulos:2019mtr}. In \cite{Harko:2018gxr} the authors introduced a coupling of the non-metricity scalar $Q$ to the matter sector, thus breaking the covariant conservation of the energy-momentum tensor, and explored the dynamical evolution for several specific coupling functions, see also \cite{Xu:2019sbp}. Models assuming the existence of a scalar field nonminimally coupled to non-metricity, called the scalar-non-metricity theories of gravity, were proposed in \cite{Jarv:2018bgs,Runkla:2018xrv}. The authors explored the resemblances of these theories with scalar-curvature and scalar-torsion models, found an equivalence of the solutions of $F(Q)$ and the metric teleparallel $F(T)$ models on the flat Friedmann cosmology, and considered conformal transformations, see also \cite{Gakis:2019rdd}.
The propagation of tensor modes in STG has been studied \cite{Conroy:2017yln,Soudi:2018dhv,Hohmann:2018xnb,Hohmann:2018wxu} exploring also the possible new 
parity-violating signatures \cite{Conroy:2019ibo,Zhao:2019xmm}. Finally, the relevance of the modified Newtonian limit in $F(Q)$ gravity to dark matter phenomenology was investigated in \cite{Milgrom:2019rtd,DAmbrosio:2020nev}.

Here we focus on a particular model of $F(Q)$ Cosmology in which the background is constructed to mimic a $\Lambda$CDM evolution in General Relativity. However, at the perturbative level, the evolution of the matter fluctuations deviates from the standard model. As it will be shown, by testing this model against redshift space distortion data, these deviations are sufficient to alleviate the present $\sigma_8$ tension. We perform a likelihood analysis in order to find the most viable parameter space, given the specific range of the dataset considered, and analyse our results.

Note that there is also a tension on the Hubble rate parameter, $H_0$, between local measurements \cite{Riess:2019cxk} and the Planck data \cite{Aghanim:2018eyx}. As this model fixes the background to exactly mimic $\Lambda$CDM, with $H_0$ being a background parameter, the model presented here does not address this problem. Note however that some studies haven shown that it is possible to solve this tension with modified $F(T)$ gravity \cite{Wang:2020zfv,Nunes:2018xbm,Basilakos:2018arq,El-Zant:2018bsc,DAgostino:2020dhv}. As there is an equivalence between $F(T)$ and $F(Q)$ at small scales \cite{Jimenez:2019ovq}, we expect it would be possible to achieve the same goal with STG.

This manuscript is organized as follows: Section \ref{csm} introduces the specific cosmology considered in this study and the equations governing the background and the evolution of matter fluctuations. On section \ref{obs} we present the dataset used in our study, expose the methodology, perform a likelihood analysis and interpret our results. Finally we conclude in section \ref{conclusions}.

\section{Cosmological model}\label{csm}

This work focuses on modified gravity models of $F(Q)$, 
characterised by the non-metricity tensor defined as
\be \label{qtensor}
Q_{\alpha\mu\nu} = \nabla_\alpha g_{\mu\nu}\,,
\ee
where the non-metricity scalar $Q$, \cite{Jimenez:2019ovq}, is the invariant constructed by the contraction,
\begin{equation}
\label{Qscalar}
Q = -Q_{\alpha\mu\nu}P^{\alpha\mu\nu}\,.
\end{equation}
The tensor $P^{\alpha\mu\nu}$ is the non-metricity conjugate,
\begin{equation} \label{conjugate}
P^{\alpha}\,_{\mu\nu} = -\frac{1}{2}L^{\alpha}{}_{\mu\nu}+\frac{1}{4}\left( Q^{\alpha}-\tilde{Q}^{\alpha} \right)g_{\mu\nu}-\frac{1}{4}\delta^{\alpha}_{(\mu}Q_{\nu )}\,,
\end{equation}
with the object
\ba
L^{\alpha}{}_{\mu\nu}   =  \frac{1}{2}Q^\alpha{}_{\mu\nu}-Q_{(\mu\nu)}{}^\alpha\,,
\ea
known as the {\it disformation}, and the two traces are 
\begin{eqnarray}
Q_{\alpha} = g^{\mu\nu}Q_{\alpha\mu\nu}\,, \hspace{1cm}
\tilde{Q}_{\alpha} = g^{\mu\nu}Q_{\mu\alpha\nu}\,.
\end{eqnarray}
We then consider the following action,
\begin{equation}
\label{action}
\mathcal{S} = \int \diff^4 x \sqrt{-g}\left[ -\frac{1}{16\pi G}F(Q)+\mathcal{L}_m(\phi,\nabla\phi) \right]\, ,
\end{equation}
where $\mathcal{L}_m$ is the Lagrangian for the matter fields. 
The field equations and geometrical interpretation on such flat and torsion-free setting where thoroughly explored in \cite{BeltranJimenez:2019tjy}.

We will now focus on a flat, homogeneous and isotropic Universe, described by the FLRW line element
\begin{equation}
ds^2 = -\diff t^2 + a(t)^2\diff {\bf x}^2 \, ,
\end{equation}
with $a(t)$ being the scale factor and $t$ represents cosmic time. The non-metricity invariant $Q$, Eq.~\eqref{Qscalar}, then becomes $Q = 6 H^2$ where $H=\dot{a}/a$ is the Hubble rate and a dot represents derivative with respect to $t$. Within this geometrical setting, the field equations emerging from the action Eq.~\eqref{action}, considering pressureless matter and a cosmological constant $\Lambda$ ($\Lambda$CDM), are \cite{BeltranJimenez:2017tkd}:
\begin{eqnarray}
2F_Q H^2 -\frac{1}{6}F &=& \frac{8\pi G}{3}\rho_m+\frac{\Lambda}{3}\, , \label{field1}\\
\left( 12H^2F_{QQ}+F_Q \right)\dot{H} &=& -4\pi G \rho_m\, ,
\end{eqnarray}
where $\rho_m$ and $\Lambda$ represent the energy density of matter and the cosmological constant, respectively, and $F_Q=\partial F/\partial Q$.

As mentioned earlier, in this current work we fill focus on the specific form of $F(Q)$ which takes us promptly to a GR background. This is achieved by setting the left hand side of Eq.~\eqref{field1} equal to $H^2$, resulting in \cite{Jimenez:2019ovq}:
\begin{equation}
\label{fq}
F(Q) = Q+ M\sqrt{Q} + C\, ,
\end{equation}
where $M$ and $C$ are constants with dimensions of mass and mass$^2$ respectively. Indeed, this choice holds the standard Friedmann equation
\begin{equation}
H^2 = \frac{8\pi G}{3}\rho_m+\frac{\Lambda}{3} + \frac{C}{6}\, ,
\end{equation}
and the matter fields then follow the continuity equation,
\begin{equation}
\dot{\rho}_m+3H\rho_m = 0 \, .
\end{equation}
For details see \cite{Jimenez:2019ovq}. Note that the constant $C$ behaves exactly as a cosmological constant. Thus, the cosmological constant term can be entirely introduced in the gravitational Lagrangian through Eq.~\eqref{fq} by setting $C=2\Lambda$ or on the matter Lagrangian fixing in this case $C=0$.

 Equation \eqref{fq} is interesting since it introduces a free parameter $M$, related to a certain mass scale, without having any influence whatsoever on the background evolution. At the perturbative level however this will modify the growth of fluctuations and may be used to fit cosmological observations, as it will be shown.

The evolution of matter overdensities, denoted by $\delta$, in the small scales limit were derived in \cite{Jimenez:2019ovq}, and can be written as,
\begin{equation}
\label{pert}
\ddot{\delta}+2H\dot{\delta} - \frac{4\pi G}{F_Q}\,\rho_m\,\delta = 0\, ,
\end{equation}
which is similar as in standard GR with a varying gravitational constant only at the perturbative level, $G_{\rm eff}/G = 1/F_Q$, sourced by the non-metricity. Note that Eq.~\eqref{pert} reduces to pure GR by setting $M=0$ in Eq.~\eqref{fq}.

\begin{figure}[b]
\begin{center}
\includegraphics[width=0.46\textwidth]{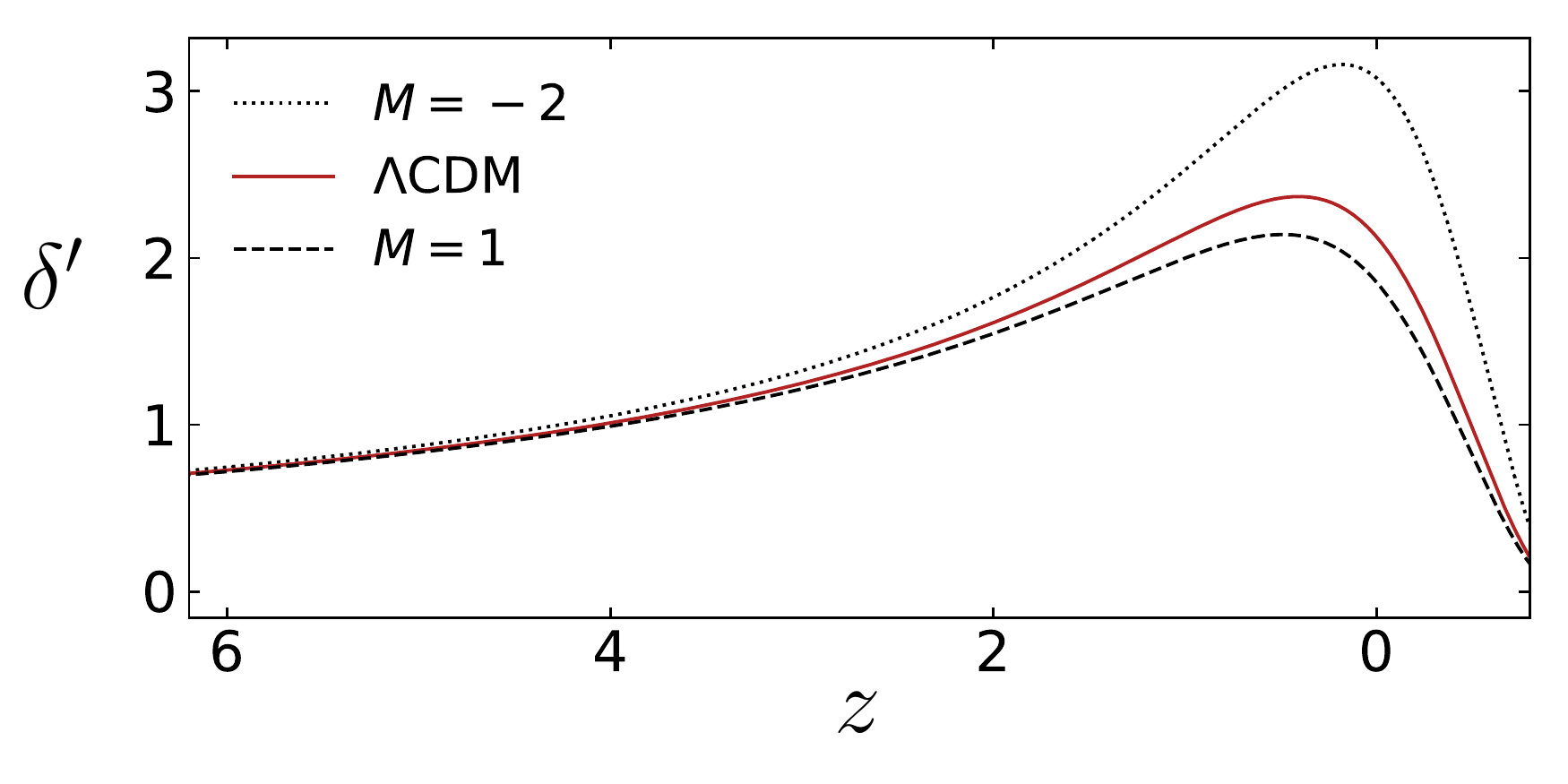}
\end{center}
\vspace{-0.5cm}
\caption{\label{m} Derivative of the density contrast, $\delta'$, for $M=-2$ (dotted), $\Lambda$CDM (solid) and $M=1$ (dashed). $M$ values are in $H_0$ units.
}
\end{figure}

It is useful to rewrite Eq.~\eqref{pert} with respect to the number of e-folds $N=\ln a$, {\it i.e.} $\dot{\delta} =H\delta'$. With $F(Q)$ as given by Eq.~\eqref{fq}, we obtain
\begin{equation}
\label{pert2}
\delta'' + \delta'\left( 2+\frac{H'}{H} \right)-\frac{3\sqrt{6}H}{2\sqrt{6}H+M}\Omega_m\delta=0  \, ,
\end{equation}
where we have defined the relative energy density parameter of matter, $\Omega_m=8\pi G \rho_m / 3H^2$. Although the background of this model remains unaffected by varying $M$, the matter perturbations are suppressed as $M$ grows, or enhanced for $M<0$, depending on the corrective force term in Eq.~\eqref{pert2}. This effect is shown in Fig.~\ref{fig1} where $\delta'$ in terms of redshift, $1+z = 1/a$, is depicted for different values of $M$. It is useful to express $M$ in units of $H_0$ and hereafter we will always present our results accordingly.

\begin{table*}[t]
\centering
\begin{tabular}{lccccccc}
\hline\hline
\\ [-0.5cm]
$\,\,${\bf Model}$\quad$ & $M$ & $\sigma_8$ & $N_{fp}$ & $\chi^2$ & $\chi^2/\,$dof & $AIC_c$ & $\Delta AIC_c$ \\ [0.01cm] \hline \\ [-0.5cm]
$\,\,\Lambda$CDM$\quad$ & {\color{gray} $0$} & $\quad0.7535\pm 0.0280\quad$ &   $1$    &   $\quad 13.1227\quad$    &  $0.6249$ & $\quad 15.3227\quad$ & $\,\, 0.5951 \,\,$\\ [0.15cm]\hline \\ [-0.5cm] 
$\,\, F(Q)= Q+ M\sqrt{Q}\quad\quad$ & $\quad 2.0331^{+ 3.8212}_{- 1.9596}\quad$ & $0.8326^{+0.1386}_{-0.0630}$ &   $2$   &   $11.9960$    &   $0.5998$ & $16.6279$ & $1.9003$    \\ [0.15cm]\hline \\ [-0.5cm]
$\,\, F(Q)= Q+\sqrt{8\Lambda Q}+2\Lambda \quad$ & {\color{gray} $4.0544$} & $ 0.8987\pm 0.0332$ &   $1$ &   $12.5276$    &  $0.5966$ & $14.7276$ & $0$    \\ 
\vspace{-0.45cm}\\ [0.09cm]
\hline
\end{tabular}
\caption{\label{tab} Best fit values for $M$ (in units of $H_0$) and $\sigma_8$, number of fitted parameters ($N_{fp}$) and respective $\chi^2$ and $AIC_c$ values.}
\end{table*}

The standard model of cosmology has survived throughout a remarkably number of observational tests so far, particularly at the background level. Therefore, and since we wish to compare $F(Q)$ predictions versus the standard $\Lambda$CDM, here we fix its parameters to the latest Planck 2018 results \cite{Aghanim:2018eyx}.

There are two specific values of $M$ of great interest that are taken as reference models. The first is a pure $\Lambda$CDM, obtained setting $M=0$, {\it i.e.} $F(Q)=Q$, with both the background and perturbations evolving as in the standard model of Cosmology.
The second value is obtained associating the mass scale $M$ with the cosmological constant ({\it i.e.} with $C$). 
Assuming that the function $F(Q)$ is of the form:
\begin{equation}
\label{cc1}
F(Q)= \left( \sqrt{Q}+\frac{M}{2} \right)^2 \, ,
\end{equation}
we obtain a cosmological constant in the Friedmann equation with $C=M^2/4$ in Eq.~\eqref{fq}. If this is to be the only cosmological constant, then $M^2 = 24H_0^2\Omega_{\Lambda}^0$, which using the latest Planck 2018 \cite{Aghanim:2018eyx} values yields $M = 4.0544$. Hence we can write this second case as,
\begin{equation}
\label{second}
F(Q)=Q+\sqrt{8\Lambda Q}+2\Lambda\, ,
\end{equation}
where the background  follows a $\Lambda$CDM evolution, but not the perturbations.

\section{Analysis and results}\label{obs}

\begin{figure}[b]
\begin{center}
\includegraphics[width=0.46\textwidth]{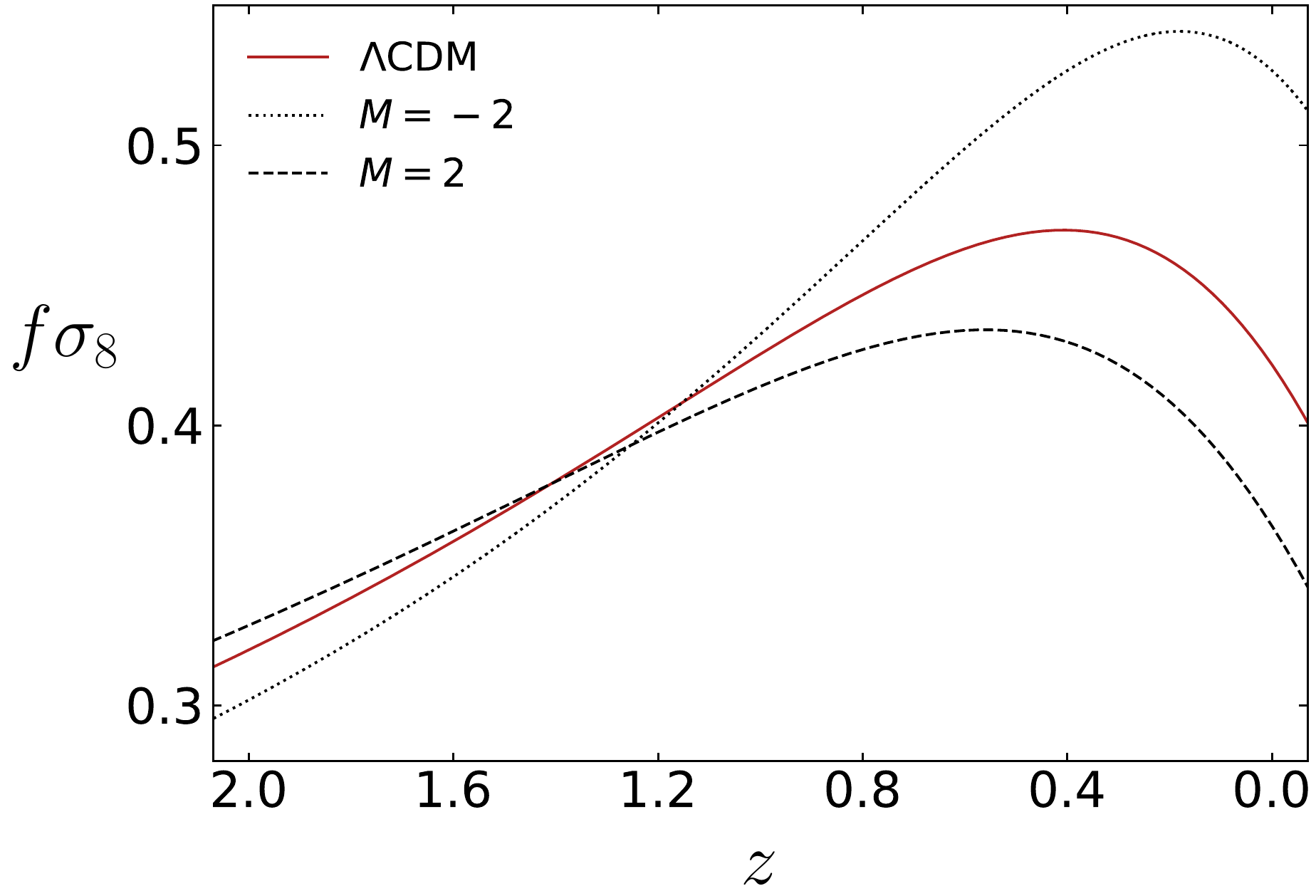}
\end{center}
\vspace{-0.5cm}
\caption{\label{m2} Evolution of $f\sigma_8$ for $M=-2$ (dotted), $\Lambda$CDM (solid) and $M=2$ (dashed) with a fixed value of $\sigma_8 = 0.8$.
}
\end{figure}

Due to the peculiar velocities of galaxies in a cluster, the shape of that cluster in redshift space appears distorted to an observer. This effect is known as redshift space distortions (RSD) \cite{Kaiser:1987qv}, and can serve as a probe of structure formation processes.

The aim of this work is to test the predictions for the growth of matter (baryons + CDM) perturbations by numerical integration of Eq.~\eqref{pert2}, against observational data of RSD. To this end, we introduce the growth rate parameter, $f = \delta'(N)/\delta(N)$, which depicts the rate at which the $\delta$ fluctuations grow. RSD data usually constrain the combination
\begin{equation}
\label{fs8}
f\sigma_8(N) = \sigma_8\frac{\delta'(N)}{\delta(0)}\, ,
\end{equation}
where $\sigma_8 = \sigma_8(0)$ is the present amplitude of the matter power spectrum at the scale of $8h^{-1}$Mpc \cite{Aghanim:2018eyx,Barros:2019hsk}. Naturally, this parameter is strongly influenced by the source term in Eq.~\eqref{pert2}, thus strongly depends on the parameter $M$. This trend is depicted in Fig.~\ref{m2}.

\subsubsection{Dataset and statistics}

In this work we use a dataset consisting of 22 data points, referenced in Table I of \cite{Sagredo:2018ahx}.

\begin{figure}[t]
\begin{center}
\includegraphics[width=0.47\textwidth]{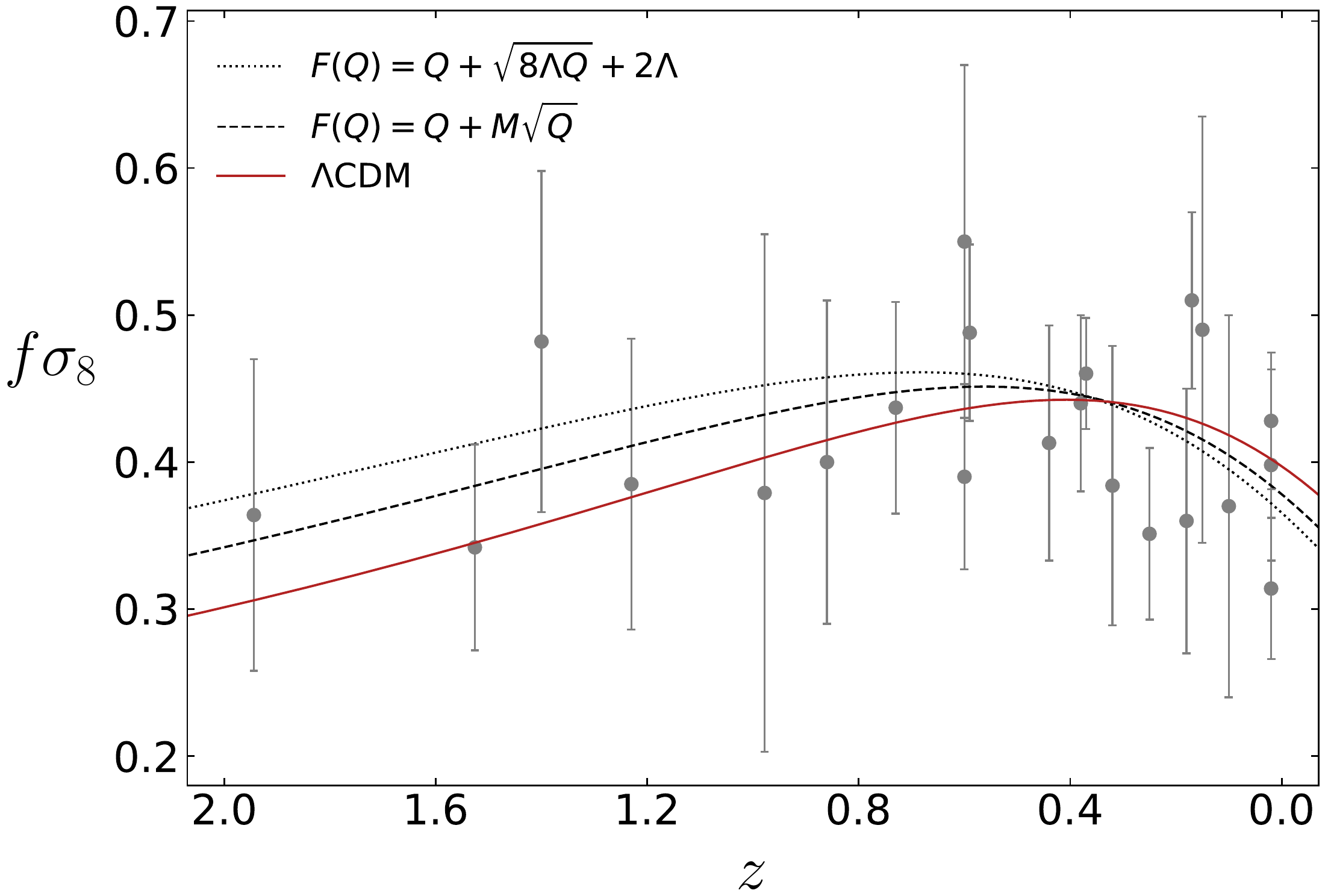}
\end{center}
\vspace{-0.6cm}
\caption{\label{fig1} Evolution of $f\sigma_8$ given the best fit values for $\Lambda$CDM (solid), $F(Q)=Q+M\sqrt{Q}$ (dashed) and $M^2=8\Lambda$ (dotted). Data points can be found on  Table I of \cite{Sagredo:2018ahx}.}
\end{figure}

The authors of \cite{Nesseris:2017vor} constructed a robust subsample of 18 independent data points, the so called 'Gold-2017' growth dataset, from a total of 34 data points for $f\sigma_8$. The validity of this dataset -- with four new data points added -- was later analysed through a Bayesian model comparison and performing cross-checks to validate its sensitivity \cite{Sagredo:2018ahx}. They confirmed that the dataset presented in Table I of \cite{Sagredo:2018ahx} is  internally robust. In this work we use these observational values for $f\sigma_8$, which have already been employed in other investigations \cite{Bouali:2019whr,Sagredo:2018rvc}. The data points and respective redshifts $z$ are shown in Fig.~\ref{fig1}.

As mentioned in \cite{Nesseris:2017vor,Sagredo:2018ahx}, the observations of the data points were conducted assuming specific values for a fiducial cosmological model in order to calculate the distances to the sources. These are also listed in Table I of \cite{Sagredo:2018ahx}. Thus, we follow the procedure described in \cite{Kazantzidis:2018rnb} correcting the $f\sigma_8$ parameter by the ratio,
\begin{equation}
\label{correction}
r(N) = \frac{H^{\rm obs} D_A^{\rm obs}}{H^{\rm th}D_A^{\rm th}}\, ,
\end{equation}
of the reference cosmology used in the observations by the theoretical model we are testing. Here, $H$ is the Hubble parameter and $D_A$ the angular diameter distance,
\begin{equation}
D_A = \frac{1}{1+z}\int_0^z\frac{1}{H(z')}dz'\, .
\end{equation}

We then perform a likelihood analysis letting $M$ and $\sigma_8=\sigma_8(0)$ to be free parameters. The likelihood is calculated through,
\begin{equation}
L = A \exp\left( -\chi^2/2 \right)\, ,
\end{equation}
where $A$ is a normalization constant and the $\chi^2$ is given by, 
\begin{equation}
\chi^2 = \left[ d_i - r(N_i)\,t_i \right]^T C_{ij}^{-1} \left[ d_j - r(N_j)\,t_j \right]\, ,
\end{equation}
taking into account Eq.~\eqref{correction}, 
with $d_i$, $t_i$ and $C_{ij}$ being the vector of data, theory and the covariance matrix. It is also useful to divide the $\chi^2$ by the degrees of freedom (dof), which equals to the number of data points, $N_d$, minus the number of fitted parameters, $N_{fp}$, resulting in the reduced $\chi^2$:
\begin{equation}
\chi^2_{\rm red} = \frac{\chi^2}{N_d - N_{fp}}\, .
\end{equation}

\begin{figure}[b]
\begin{center}
\includegraphics[width=0.48\textwidth]{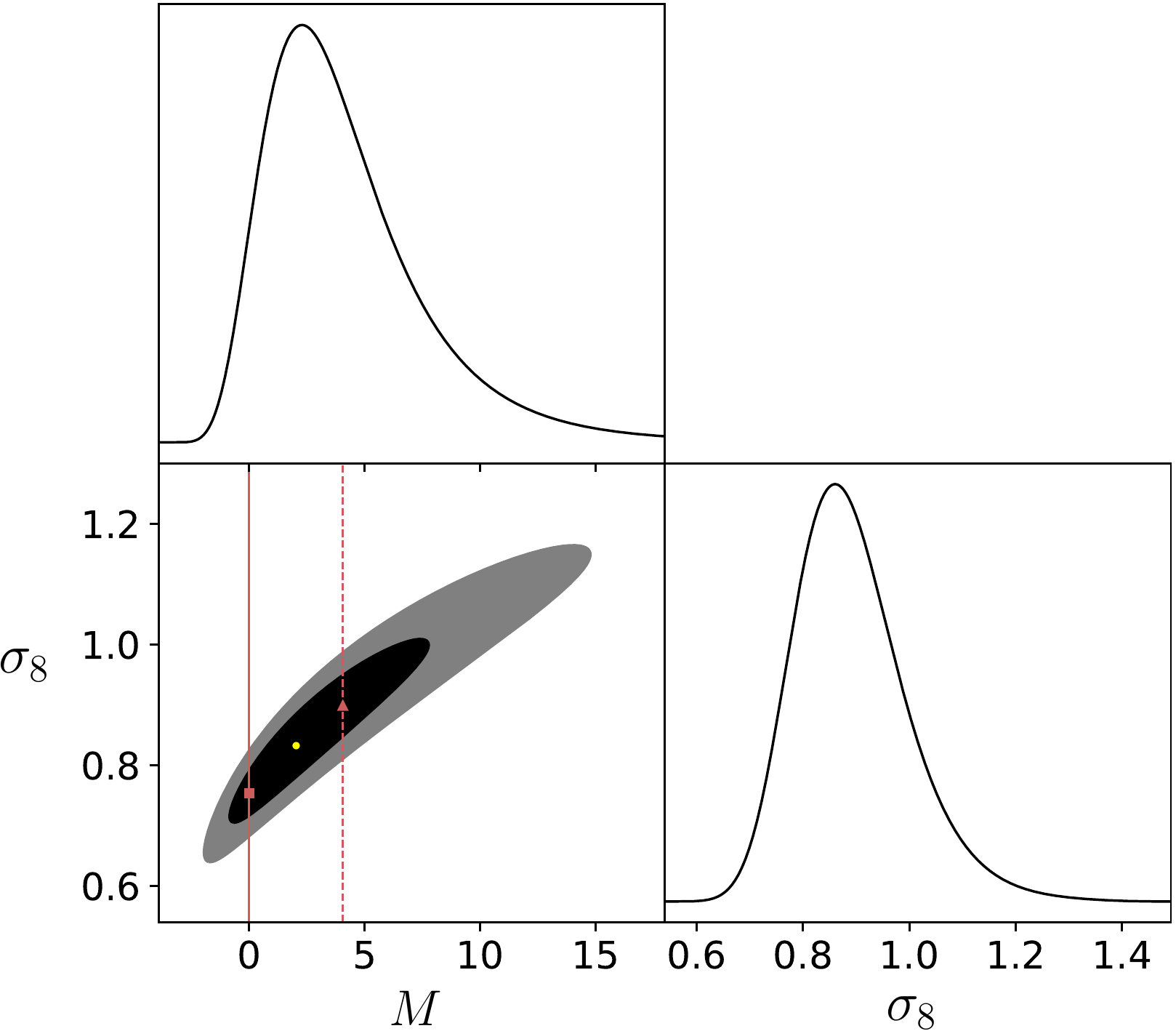}
\end{center}
\vspace{-0.5cm}
\caption{\label{fig2} Constraints on $M$ and $\sigma_8$: contours for the 1$\sigma$ and 2$\sigma$ regions and the respective marginalized curves for $F(Q)=Q+M\sqrt{Q}$. A circle denotes the best fit value. The vertical lines denote the reference models: $\Lambda$CDM (solid) with a square marker on the best fit and for $M^2=8\Lambda$ (dashed) with a triangle marker on the best fit.}
\end{figure}

Comparing the reduced $\chi^2$ gives us a rough estimate on how a model is preferred among others. However, there 
are more rigorous criteria available for model comparison. Among them, a widely used test in the literature is the Akaike Information Criterion ($AIC$) \cite{Akaike1974}. The $AIC$ formula \cite{Bouali:2019whr,Sagredo:2018rvc},
\begin{equation}
AIC = -2\ln \left( L_{\rm max} \right) + 2N_{fp}\, ,
\end{equation}
takes into account the number of fitted parameters, and in the case of a small sample size can be corrected as
\begin{equation}
AIC_c = AIC + 2\frac{N_{fp}\left( N_{fp} +1 \right)}{N_d - N_{fp}-1}\, ,
\end{equation}
in order to avoid over-fitting of the data. Note that the correction becomes superfluous for $N_d\rightarrow\infty$. Naturally, the model with smaller $AIC_c$ is favoured. However, we will focus on the deviations $\Delta AIC_c = AIC_c - AIC_c^{\rm min}$ from the model which minimizes the $AIC_c$.
Models with $\Delta AIC_c\leq 2$, $4\leq\Delta AIC_c\leq 7$ and $\Delta AIC_c\geq 10$ have substantial support, considerably less support and essentially no support, respectively \cite{10.1177/0049124104268644}. 

\begin{figure*}[t]
\begin{center}
\includegraphics[width=0.85\textwidth]{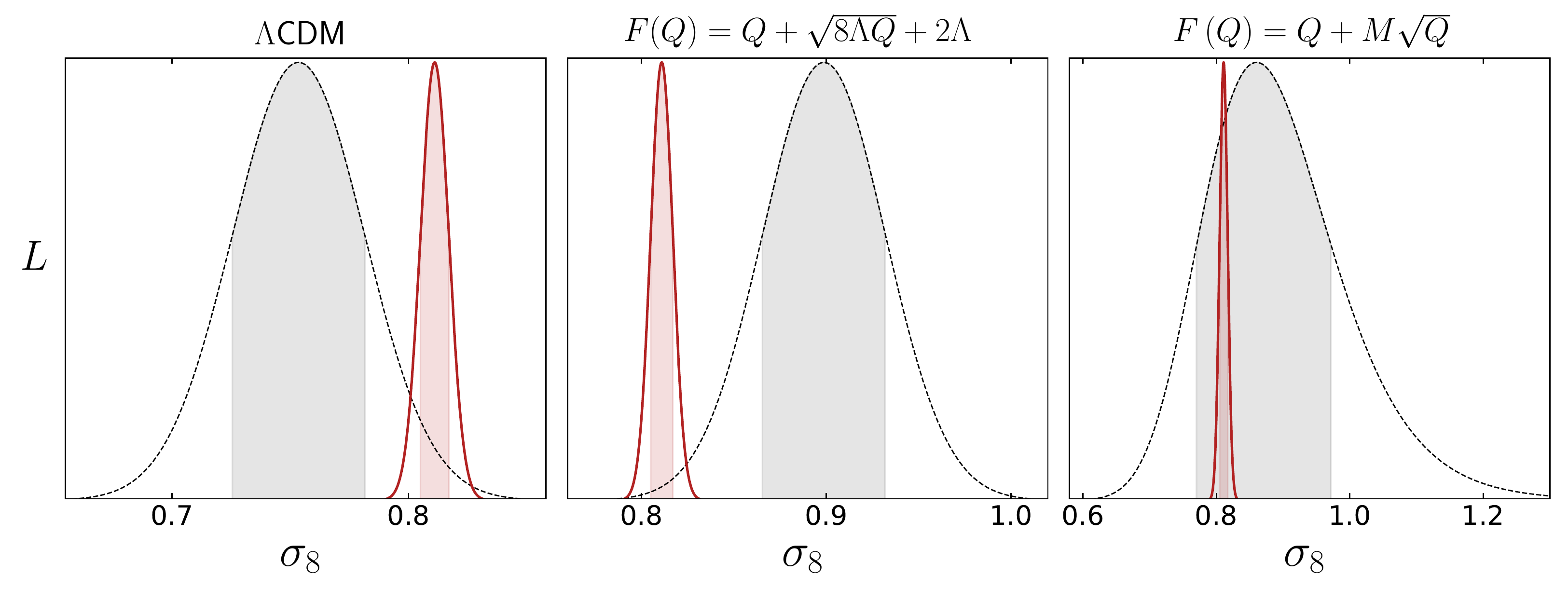}
\end{center}
\vspace{-0.5cm}
\caption{\label{ten} Likelihood for $\sigma_8$ (curves) and respective $1\sigma$ interval (shaded region), for the models labeled on top (dashed lines), and the Planck reference (solid lines).
}
\end{figure*}

\subsubsection{Results}

The best fit values are presented on Table.~\ref{tab}, with the respective statistical uncertainties. The evolution of $f\sigma_8$ for the best fit values encountered for the three models discussed is illustrated in Fig.~\ref{fig2}. The best fit for $\Lambda$CDM is $\sigma_8=0.7535\pm 0.0280$ which is known to be in tension with the Planck 2018 values $\sigma_8=0.811\pm 0.006$ \cite{Aghanim:2018eyx}. 
On the other hand, introducing one more parameter, $M$, the $F(Q)$ cosmology has a best fit $M=2.0331^{+ 3.8212}_{- 1.9596}\neq 0$, which curiously does not include $\Lambda$CDM ($M=0$) at the level of $1\sigma$, and $\sigma_8= 0.8326^{+0.1386}_{-0.0630}$. 
The $F(Q)=Q+\sqrt{8\Lambda Q}+2\Lambda$ model has a best fit of $\sigma_8=0.8987\pm 0.0332$.
This interval for $\sigma_8$ is also in tension with Planck, though in the opposite side of the Planck value.  
The contour regions and marginalized curves are depicted in Fig.~\ref{fig2} for $F(Q)=Q+M\sqrt{Q}$, together with the best fit values for the two reference models, $\Lambda$CDM (solid vertical line) and $F(Q)=Q+\sqrt{8\Lambda Q}+2\Lambda$ (dashed vertical line). The $\sigma_8$ value found has relatively larger uncertainty due to its degeneracy with $M$, and includes the Planck 2018 best fit well within the marginalized $1\sigma$ constrains, suggesting that it is possible to solve this tension within this modified gravity framework. The likelihoods for $\sigma_8$ regarding the three models are depicted in Fig.~\ref{ten} with the Planck likelihood (approximated as a Gaussian) for reference. Both the $\chi^2$ and $AIC_c$ tests favour the $F(Q)=Q+\sqrt{8\Lambda Q}+2\Lambda$ model which has a best fit of $\sigma_8=0.8987\pm 0.0332$, with only one fitted parameter. Of course the values for the $AIC_c$ are all relatively close, hence there is no strong evidence for a preferred model.

\section{Conclusions}\label{conclusions}

In this work we have numerically evolved the linear matter perturbations for two models of $F(Q)$ cosmology, and tested them against redshift space distortions data.
The first model adds one more parameter to the standard $\Lambda$CDM, and from the value of the $\chi^2_{\rm red}$ parameter of the statistical analysis, turns out to be a better fit to the data. The central value of $\sigma_8$ is very close to the value obtained with Planck, within the uncertainty of the analysis, thus alleviating the present $\sigma_8$ tension between RSD and the Planck data.  
The second model associates the cosmological constant entirely to the gravitational Lagrangian through a suitable choice of the mass scale $M$ in Eq.~\eqref{fq}. This case appears to be favoured over $\Lambda$CDM judging from the $\chi^2_{\rm red}$ and $AIC_c$ test values, notwithstanding a tension with Planck as the value of $\sigma_8$ is far larger than the one obtained with Planck. 

This work explicitly shows how the presence of a free parameter, affecting only first and higher order perturbations, in modified gravity frameworks, in particular $Q$-gravity, may naturally be tailored to fit cosmological data and alleviate tensions on perturbative observables. 
The theory seems promising and deserves a complete analysis with several observables.

\acknowledgements{
This research was supported by Funda\c{c}\~ao para a Ci\^encia e a Tecnologia (FCT) through the research grants: UID/FIS/04434/2019, PTDC/FIS-OUT/29048/2017 (DarkRipple),  COMPETE2020: POCI-01-0145-FEDER-028987 \& FCT: PTDC/FIS-AST/28987/2017 (CosmoESPRESSO) and IF/00852/2015 (Dark Couplings). B.J.B is supported by the grant PD/BD/128018/2016 (PhD::SPACE program) and N.J.N by the contract and exploratory project IF/00852/2015 
(Dark Couplings) from Funda\c{c}\~ao para a Ci\^encia e Tecnologia. 
TK is supported by the Estonian Research Council through the Personal Research Funding project PRG356 (Gauge Gravity) and by the European Regional Development Fund through the Center of Excellence TK133 (The
Dark Side of the Universe). This article is based upon work from COST Action CA15117 (CANTATA), supported by COST (European Cooperation in Science and Technology).}

\bibliography{bib}

\end{document}